\documentclass[aps,reprint,preprintnumbers,twocolumn,floatfix,superscriptaddress,prl,showpacs,longbibliography]{revtex4-1}
\usepackage{physics}
\usepackage{natbib}
\usepackage{amsmath,amssymb}
\usepackage{graphicx}
\usepackage[dvipsnames]{xcolor}
\usepackage[caption=false]{subfig}
\usepackage{float}
\usepackage[pdftex]{hyperref}

\begin{document}
	\title{Optical Interferometry with Quantum Networks}
	\author{E. T. Khabiboulline}
	\email{ekhabiboulline@g.harvard.edu}
	\affiliation{Department of Physics, Harvard University, Cambridge, Massachusetts 02138, USA}
	\author{J. Borregaard}
	\affiliation{Department of Physics, Harvard University, Cambridge, Massachusetts 02138, USA}
	\affiliation{QMATH, Department of Mathematical Sciences, University of Copenhagen, 2100 Copenhagen \O, Denmark}
	\author{K. \surname{De Greve}}
	\author{M. D. Lukin}
	\affiliation{Department of Physics, Harvard University, Cambridge, Massachusetts 02138, USA}
	
	\begin{abstract}
	We propose a method for optical interferometry in telescope arrays assisted by quantum networks. In our approach, the quantum state of incoming photons along with an arrival time index is stored in a binary qubit code at each receiver. Nonlocal retrieval of the quantum state via entanglement-assisted parity checks at the expected photon arrival rate allows for direct extraction of the phase difference, effectively circumventing transmission losses between nodes. Compared to prior proposals, our scheme (based on efficient quantum data compression) offers an exponential decrease in required entanglement bandwidth. Experimental implementation is then feasible with near-term technology, enabling optical imaging of astronomical objects akin to well-established radio interferometers and pushing resolution beyond what is practically achievable classically.
	\end{abstract}
	
	\maketitle

	High-resolution imaging using large telescope arrays is by now a well-established technique in the microwave and radio-frequency domains~\citep{Thompson1999,kellermann2001}. Although extending to the optical domain may offer substantial advantages in terms of resolution~\citep{Stee2017,brummelaar2005}, this task is extremely challenging in practice. The requirement of interferometric stabilization at optical wavelengths and the weakness of light sources in this domain have precluded the widespread adoption of optical telescope arrays~\citep{Thompson2007}. Notably, the weaker light intensities make phase-sensitive heterodyne detection infeasible due to vacuum fluctuations~\citep{Tsang2011}; therefore, high-resolution optical telescopes are operated by directly interfering the collected light~\citep{darre2016}. Then, the size of the array (and consequently resolution) is ultimately limited by transmission losses between telescope sites. 

	In this Letter, we propose a new approach to overcome these limitations with networks~\citep{Kimble2008} of quantum memories connected via entanglement. Specifically, we describe a scheme for efficiently determining the optical phase difference between two widely separated receivers. Each detector runs a ``quantum shift register'' storing incident photon states at a rate that is matched to the inverse detection bandwidth. Then, at the anticipated mean photon arrival rate, the memories are interrogated with entangled pairs to provide information akin to that obtained from a radio interferometer. Employing quantum repeater techniques~\citep{Briegel1998}, this approach completely circumvents transmission losses. The resulting increase in baseline to arbitrarily large distances potentially allows for substantial enhancement in imaging resolution~\citep{brummelaar2005}.
	
	\begin{figure} [t]
		\centering
		\includegraphics[width=0.5\textwidth]{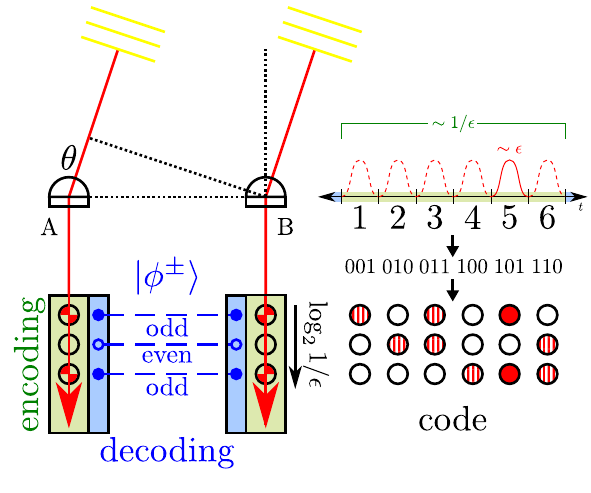}
		\caption{Overview of basic operation. Light from a distant source is collected at two sites and stored in quantum memory over time bins digitized by detector bandwidth. Both the quantum state and the arrival time of an incident photon are encoded in a binary qubit code. For example, if the photon arrives in the fifth time bin, corresponding to binary representation 101, we store it in a quantum state with flipped first and third qubits at each node. Decoding of the arrival time is accomplished by nonlocal parity checks assisted by entangled pairs, projecting the memories onto a known entangled state. The phase information can then be extracted without directly interfering the signal from the two memories, thus circumventing transmission losses. Network resources scale only logarithmically with source intensity $\epsilon$.
			\label{fig:memoryScheme}}
	\end{figure}
	
	Before proceeding, we note that the use of entanglement to connect remote telescope sites has been proposed previously by means of postselected quantum teleportation of incident optical photons~\citep{Gottesman2012}. The key limitation of this visionary proposal is the requirement of an excessive amount of distributed entangled pairs. They must be supplied at a rate similar to the spectral bandwidth of the optical telescope, which is currently not feasible. In the relevant case of weak sources, such that incident photons are rare, the use of quantum memories circumvents this requirement. Under distributed compression, the incoming light is efficiently processed using only $\sim$$\log_2(1/\epsilon)$ memory qubits and entangled pairs, where $\epsilon\ll1$ is the mean photon number. Then, the entire loss-free interferometric operation can be realized with modest quantum nodes consisting of about 20 qubits and a distributed entanglement rate in the 200 kHz range. Our proposal realizes an effective event-ready scheme, avoiding the wasteful expenditure of entanglement for vacuum events.
	
	In our protocol, illustrated in Fig.~\ref{fig:memoryScheme}, incoming light is stored by flipping stationary quantum bit memories~\citep{ritter2012} at each telescope site. The storage procedure operates over a time interval set by the detector bandwidth. Multiple qubits are needed to record a photon spread over many time bins. We assume that the light is weak such that most time bins contain vacuum. Consider first a unary encoding with one memory qubit for each time bin. After the single photon is stored, the memories are in a superposition between one site having the excitation versus the other: e.g., $(\ket{000010\ldots}_A\ket{000000\ldots}_B+e^{i\theta}\ket{000000\ldots}_A\ket{000010\ldots}_B)/\sqrt{2}$, where the memory register corresponds to the time bin. The goal of the interferometer is to extract the relative phase $\theta$. To determine which memory qubits to use for interferometry without collapsing the superposition, the parity of parallel memory registers can be checked with an entangled state per register. Introducing Bell pairs ($\ket{\phi^{\pm}}=(\ket{0,0}\pm\ket{1,1})/\sqrt{2}$, where $\ket{i}_A\ket{j}_B\equiv\ket{i,j}$), controlled phase (CZ) gates between the memory qubits on either side and the entangled pairs have the following effect:
	\begin{eqnarray}
	(\ket{0,0},\ket{1,1})\ket{\phi^+}&\xrightarrow{2 \times \text{CZ}}&(\ket{0,0},\ket{1,1})\ket{\phi^+}\,, \\
	(\ket{0,1},\ket{1,0})\ket{\phi^+}&\xrightarrow{2 \times \text{CZ}}&(\ket{0,1},\ket{1,0})\ket{\phi^-}\,.
	\end{eqnarray}
	A measurement in the $X$ basis (projecting on the states $\ket{\pm}=(\ket{0}\pm\ket{1})/\sqrt{2}$) of each qubit in the Bell pair then reveals their parity, from which we can infer arrival time because the odd-parity register is the one containing the excitation. Its relative phase can be subsequently extracted, e.g., via measurement of one of the qubits in the $X$ basis and the other in a rotated basis to interfere the phase. Similar to a prior scheme~\citep{Gottesman2012}, the unary code requires one entangled pair for each time bin, implying large consumption for practical bandwidth.
	
	Encoding in binary accomplishes the same task but with a logarithmic scaling of resources. We exploit that only one photon arrives over $M\sim1/\epsilon$ time bins; i.e., we only have to store one logical qubit per block. Label each time bin $m\in\mathbb{Z_+}$ with its binary representation $m_2$, and define logical qubits $\ket{\bar{0}}\equiv\ket{0\ldots0}$ and $\ket{\bar{1}_m}\equiv\ket{m_2}$. For example, the fifth time bin is encoded as $\ket{\bar{1}_5}=\ket{1010...0}$, which is formed from physical qubits at one site. Generally, $\log_2(M+1)$ bits are needed to losslessly encode $M$ possible arrival times plus the vacuum. This encoding is performed by a logical controlled not (CX) gate, which is a product of physical CX gates between the control photonic qubit and target memory qubits specified by the binary representation:
	\begin{eqnarray}
	\ket{0}(\ket{\bar{0}},\ket{\bar{1}_j})&\xrightarrow{\overline{\text{CX}}_m}&\ket{0}(\ket{\bar{0}},\ket{\bar{1}_j})\,, \\
	\ket{1}(\ket{\bar{0}},\ket{\bar{1}_j})&\xrightarrow{\overline{\text{CX}}_m}&\ket{1}(\ket{\bar{1}_m},\ket{\bar{1}_j+\bar{1}_m})\,.
	\end{eqnarray}
	The encoding keeps track of the arrival time of one photon: empty time bins leave the memories unchanged, whereas only the time bin that does contain a photon maps into the quantum memory via the binary code. Decoupling the photonic qubit through an $X$-basis measurement completes the encoding step but imparts a conditional phase associated with each time bin~\citep{SM}. This phase must be corrected, requiring knowledge of the time bin when the photon did arrive. The arrival time can be decoded while preserving spatial coherence by applying nonlocal parity checks on qubit pairs in the same register analogously to the case of unary encoding described earlier. For example, if the photon arrived in the fifth time bin, the Bell pairs would be found in the state $\ket{\phi^-}\ket{\phi^+}\ket{\phi^-}\ket{\phi^+}\ldots\ket{\phi^+}$. Because there are $\log_2(M+1)$ qubits per memory, $\log_2(M+1)$ preestablished entangled pairs are also consumed.
	
	Besides identifying the photon arrival time, the parity checks project out the vacuum component of the state. Modeling the astronomical object as a weak thermal source~\citep{Zmuidzinas2003,Tsang2016}, the light arriving in each time bin is described by a density matrix~\citep{Tsang2011}
	\begin{eqnarray} \label{eq:rhoAB}
	\rho_{AB}&=&(1-\epsilon)\rho_{\text{vac}}+\frac{\epsilon(1+\abs{g})}{2}\ket{\psi^{+}_{\theta}}\bra{\psi^+_{\theta}} \nonumber \\
	&&+\frac{\epsilon(1-\abs{g})}{2}\ket{\psi^{-}_{\theta}}\bra{\psi^-_{\theta}}+O(\epsilon^2)\,,
	\end{eqnarray}
	to the first order in $\epsilon$, where $\ket{\psi^{\pm}_{\theta}}=(\ket{0,1}\pm e^{i\theta}\ket{1,0})/\sqrt{2}$ and $\rho_{\text{vac}}=\ket{0,0}\bra{0,0}$ in the photon-number basis. The first-order spatial coherence $g=\abs{g}e^{i\theta}$, also known as the visibility, generally has amplitude $\abs{g}\leq1$. The nonlocal parity checks project onto the logical qubit states. After acting on $M\sim1/\epsilon$ samples of $\rho_{AB}$, the memories likely contain one logical excitation. This postselection via measurement leads to efficient error accumulation, as elaborated on below. The visibility $g$ can then be extracted through a logical measurement similar to the case of unary encoding discussed above~\citep{SM}.
	
	Specifically, for the example of a photon being detected in the fifth time bin, the memory ends up in the following entangled state up to a known phase flip from the state transfer operation:
	\begin{equation}
	\frac{(1\pm\abs{g})}{2}\ket{\bar{\psi}_{\theta}^+}\bra{\bar{\psi}^+_{\theta}}+\frac{(1\mp\abs{g})}{2}\ket{\bar{\psi}^-_{\theta}}\bra{\bar{\psi}^-_{\theta}}\,,
	\end{equation}
	where $\ket{\bar{\psi}^{\pm}_\theta}=(\ket{\bar{0},\bar{1}_5}\pm e^{i\theta}\ket{\bar{1}_5,\bar{0}})/\sqrt{2}$, which has four entangled physical qubits $(\ket{00,11}\pm e^{i\theta}\ket{11,00})/\sqrt{2}$. The other, even-parity qubits are in the $\ket{0}$ state and can be traced out. After measuring the first three of the four entangled qubits in the $X$ basis, the remaining qubit is in the state
	\begin{equation}
	\frac{1}{2}\left(\ket{0}\bra{0}+\ket{1}\bra{1}+(-1)^{n_{-}}(g\ket{0}\bra{1}+h.c.)\right)\,,
	\end{equation}
	where $n_-$ is the number of $\ket{-}$ outcomes from the $X$-basis measurements. Assume $n_-=0$ for simplicity. Applying a phase shift $U_\delta=\ket{0}\bra{0}+e^{i\delta}\ket{1}\bra{1}$ and measuring the qubit in the $X$ basis will have outcome $\ket{\pm}$ with probability $(1\pm\Re(ge^{-i\delta}))/2$. The visibility $g$ can then be extracted by sweeping over $\delta$ and repeating the above procedure for many photons.

	Physically, our scheme can be implemented with long-lived atomic ground states (such as in Rb atoms)~\citep{specht2011}, solid-state qubits with an optical interface (such as SiV defect centers in diamond)~\citep{sivkurtsiefer,sivrogers,sivalp}, or spin qubits in quantum dots~\citep{Press2008}, which have fast control. Optical cavities ensure strong light-matter interaction~\citep{sivcqed}, which is potentially matched via quantum frequency conversion~\citep{pra}. Absorption of the photon by an auxiliary atom in a Raman setup~\citep{cirac1997,boozer2007} enables easy $X$-basis measurement, with the same atom reused for every optical mode. Maintaining a stable phase at the level specified by detector bandwidth can be accomplished with a reference laser, as done in atomic clocks~\citep{clockreview}. Logical CX gates between the auxiliary atom and the memory atoms could be realized as cavity-mediated~\citep{pellizzari1995} or Rydberg gates~\citep{saffman2010}. Alternative schemes involve photon-atom gates~\citep{tiecke2014,ritter2012} and photon detection, eliminating the auxiliary atom~\citep{pra}.
		
	The dominant errors in our protocol will arguably originate from the two-qubit gates. Note, however, that all time bins except the one containing the photon will result in trivial CX gates where the control qubit is in state $\ket{0}$. The trivial action of the CX should have an error rate less than $\epsilon(\log_2 1/\epsilon)^{-1}$ to preserve the memory. A number of gate schemes satisfy this criterion~\citep{SM}; for example, in photon-atom gates, the absence of a photon does not affect the atom. We therefore assume that only the nontrivial CX operations lead to significant errors. We also consider higher-order corrections to the photonic density matrix in Eq.~(\ref{eq:rhoAB}), which introduce multiple-photon events leading to undetectable errors in the binary code. 
	
	\begin{figure} [t]
	\centering \includegraphics[width=0.5\textwidth]{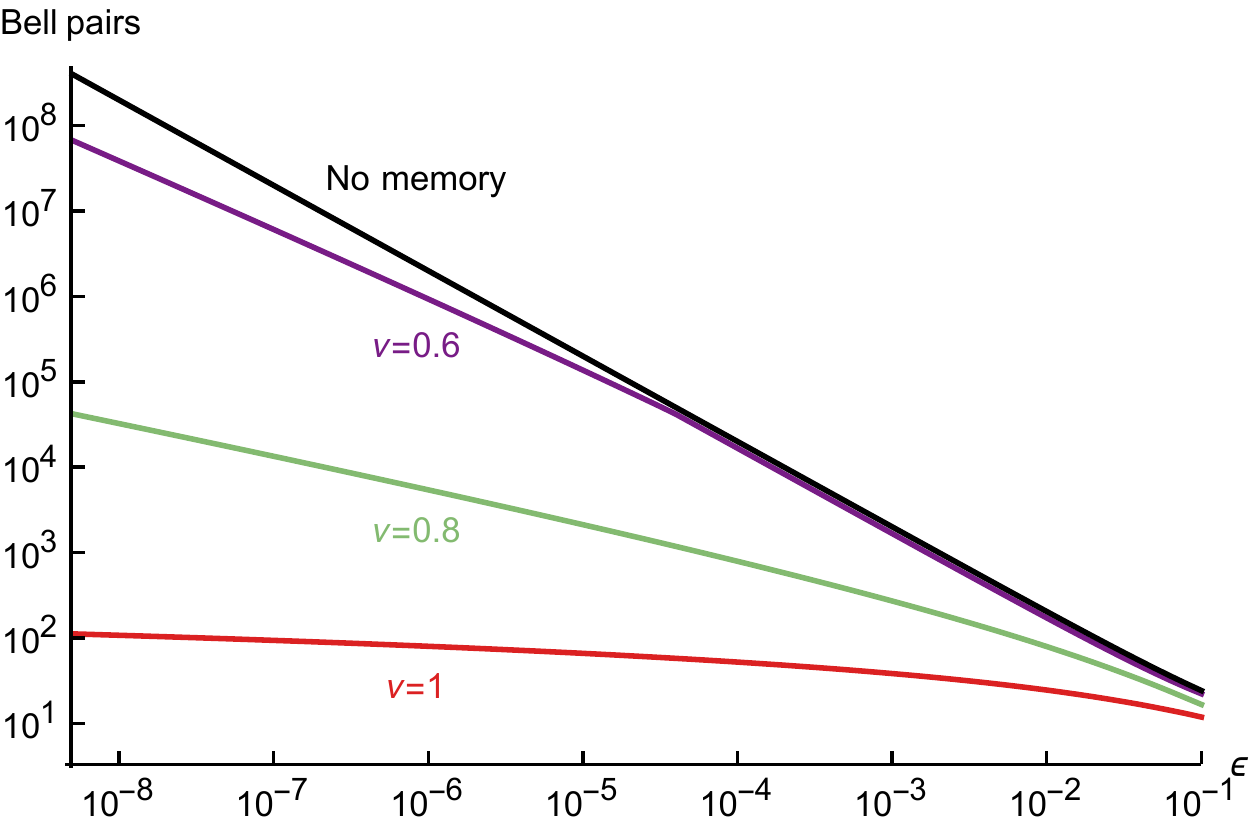}
	\caption{Minimum number of Bell pairs needed to attain $\lVert F\rVert\geq1$, corresponding to variance $\lesssim1$ in estimating the visibility $g$, as a function of source intensity $\epsilon$. We consider errors in the coding operations parametrized by decreasing $\nu$, as detailed in the main text. Ideally ($\nu=1$), $M\sim1/\epsilon$ time bins are encoded per block and read out with $\log_2(M+1)$ entangled pairs. For sufficiently large errors ($\nu=0.6$), the encoding fails and memoryless operation with the readout of every time bin is recovered, similar to the scheme of Ref.~\citep{Gottesman2012}.
		\label{fig:fisher}}
	\end{figure} 
	
	The performance of our scheme is analyzed using the Fisher information, quantifying how much information about a given parameter can be extracted through measurement on a quantum state~\citep{Braunstein1994}. In our case, we wish to estimate the visibility $g=g_1+i g_2$, where $g_1$ and $g_2$ are two real parameters, and so the Fisher information becomes a matrix. Taking the trace norm, $\lVert F\rVert$ quantifies the total obtainable information about $g$: $1/\lVert F\rVert$ has the operational interpretation of constraining the variance of the measured data~\footnote{The Cr\'amer-Rao bound~\citep{Helstrom1969} is $\Sigma\geq1/F$, where $\Sigma$ is the covariance matrix of the parameters being estimated. Note that $\lVert\Sigma\rVert\geq1/\lVert F\rVert$ and the eigenvalues of $\Sigma$ are the variances along the principal components of the measured data.}. Ideal nonlocal and local schemes operating on the state of Eq.~(\ref{eq:rhoAB}) are separated by $\lVert F\rVert\geq M\epsilon$ and $\lVert F\rVert\leq M\epsilon^2$, respectively~\citep{Tsang2011}. Intuitively, the nonlocal bound corresponds to the probability of detecting a photon from a source of intensity $\epsilon$, whereas the local bound is a factor of $\epsilon$ worse due to the inability to discriminate against the vacuum component of the state. Thus, the Fisher information explicitly demonstrates how nonlocal schemes like ours are superior to local schemes like heterodyne detection for measuring weak thermal light, as found in the optical domain.
	
	Incorporating errors and operating over $M$ time bins, we find the following norm of the Fisher information, which is bounded from below~\citep{SM}:
	\begin{equation}
	\lVert F\rVert\geq\frac{(M\epsilon)^2}{[(1+\epsilon)^M-1](1+\epsilon)^{M+2}} \mu^2  \nu^{\log_2(M+1)}\equiv \lVert F\rVert_{\text{min}}\,,
	\label{eq:F}
	\end{equation}
	where $\mu\equiv p_t(2f_t-1)^2(2f_1-1)^2$ contains the error in mapping the photonic state to memory at one site, in terms of the success probability $p_t$ and fidelity $f_t$ of the light-memory transfer and the fidelity of the one-qubit measurement $f_1$, and $\nu\equiv(2f_1-1)^4(2f_2-1)^4(2f_e-1)$ contains the error of memory coding and readout in terms of the fidelities of one-qubit measurements $f_1$, nontrivial two-qubit gates $f_2$, and preshared entanglement $f_e$. The success probability includes photon loss, which effectively increases the vacuum component of the state. Other errors are not detectable and are modeled as depolarizing channels as a worst-case scenario. Note that $\lVert F\rVert_{\text{min}}\sim M\epsilon$ for small $\epsilon$, as expected for a nonlocal scheme. Larger $M$ improves the probability of receiving the signal of a single photon, and hence $\lVert F\rVert_{\text{min}}$ initially increases; eventually, however, a maximum is reached due to competition from multiple-photon events and imperfect quantum operations. Therefore, the scheme can be optimized with respect to $M$, operating with the minimum entanglement expenditure needed to extract the information content of one photon: $\lVert F\rVert_{\text{min}}\geq1$ (see Fig.~\ref{fig:fisher}). In the ideal case ($\nu=1$), the number of entangled pairs is logarithmic in $1/\epsilon$, which is the number of time bins operated over so that roughly one photon arrives, on average. Information-theoretic arguments, based on the conditional entropy of the state described by Eq.~(\ref{eq:rhoAB}), predict a bound with the same scaling, within a constant prefactor~\citep{SM}. When errors are sufficiently large ($\nu\approx 60\%$), the memories are read out after a single time bin. Thus, an effective memoryless scheme similar to that of Ref.~\citep{Gottesman2012} is recovered, with entanglement consumption scaling as $1/\epsilon$~\footnote{A priori, there is a factor of 2 improvement from not having to postselect out half the photon events that do not reveal which-path information; our scheme always preserves coherence between sites. However, this advantage is counterbalanced by multiple-photon and gate errors}.
	
	Assuming an effective detection bandwidth $\delta_f=10$ GHz, a total area of photon collection of 10 m$^2$, and imaging in the $V$ band (centered around 555 nm), we can estimate the resources needed for a star of magnitude 10 (corresponding to $\epsilon=7\cross10^{-7}$), which is around the limit of the Center for High Angular Resolution Astronomy (CHARA) optical interferometric array~\citep{brummelaar2013}. Our ideal, optimized scheme requires $\log_2 1/\epsilon\sim20$ memory qubits per site and an entanglement distribution rate of $\delta_f \epsilon \log_2 1/\epsilon\sim200 \text{\ kHz}$~\footnote{Due to buildup of error, the optimized block length $1/\epsilon$ of encoded or decoded time bins decreases by a factor of $(0.7,0.4,0.1)$ for $\nu\in(1,0.8,0.6)$. Also note that accounting for storing the preshared entangled pairs gives another factor of 2 for the memory requirement}. The improvement over the rate necessary for a memoryless scheme~\citep{Gottesman2012} with the $10$ GHz effective bandwidth is a factor of $5\cross10^4$. Extending the current limit of the 330 m baseline of CHARA~\citep{Pedretti2009} to realistic quantum network scales greater than 10 km would increase resolution from the milliarcsecond to the microarcsecond regime~\citep{SM}. Finally, we note that the bandwidth of direct interferometers can, in practice, be adjusted to enhance signal strength at the expense of resolution~\cite{Pedretti2009}. Although the present method is limited by the bandwidth of quantum memory, our approach readily extends to broadband operation, as discussed in Ref.~\cite{pra}.
	
	Turning to possible implementations of these ideas, we note that the chosen detector bandwidth (10 GHz) sets the timescale of the encoding operation and the photon arrival rate (1 kHz) determines the memory coherence time. In Table~\ref{tab:specifications} of the Supplemental Material~\citep{SM}, we compare the capabilities of various physical realizations. Among the most promising candidates are SiV centers in diamond, striking a balance with gate time on the order of nanoseconds and millisecond-scale coherence. Techniques such as parallelization, repeated readout, and photon detection~\citep{pra} improve performance with a modest overhead in resources. Furthermore, we emphasize that our scheme performs well in the presence of noise, which ultimately reduces the interference, as seen in Fig.~\ref{fig:fisher}. Therefore, it is amenable to experimental testing and development of noisy intermediate-scale quantum devices. Demonstrating kilohertz-scale entanglement generation between remote quantum memories, high-fidelity light-matter interfaces, and the addressing of multiple qubits at network nodes would facilitate the realization of our proposal.
	
	\begin{figure} [t]
		\centering \includegraphics[width=0.5\textwidth]{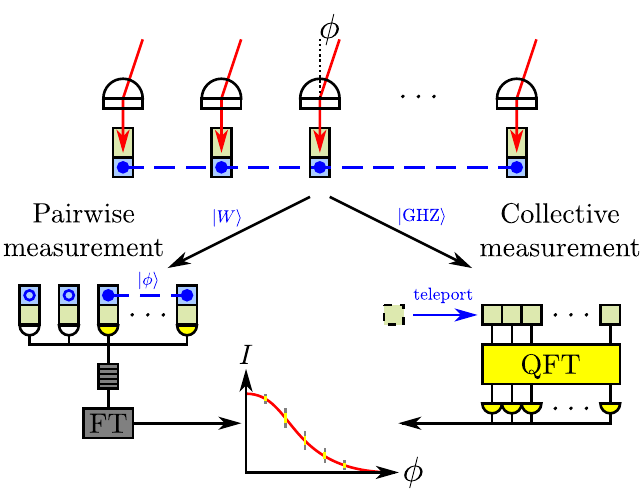}
		\caption{Generalization to $N>2$ sites in the telescope array. Decoding with a $W$ state collapses the network state to two nodes, and the protocol continues as before. The visibility data are stored in a classical memory until enough events have accumulated to perform a Fourier transform (FT). Using GHZ states instead preserves coherence across the network. Quantum teleporting the memories to one site for convenience, a quantum Fourier transform (QFT) is applied, yielding the desired intensity distribution directly as the probabilities of measurement outcomes.
			\label{fig:network}}
	\end{figure}
	
	In order to image a broad object, telescope arrays consisting of $N>2$ sites are used to sample the visibility $g(x)$ across $N-1$ points between $x=0$ and $x=b$ ($b$ is the baseline, or maximum length). According to the van Cittert-Zernike theorem~\citep{Zernike1938}, a Fourier transform yields an estimate of the stellar intensity distribution $I(\phi)$. To operate in this manner, our network protocol generalizes to $N>2$ nodes. Under conditions when one photon is incident on the telescope array, we encode the optical modes in a binary code, as in the two-node case. The nonlocal parity checks are performed using either $N$-qubit Greenberger-Horne-Zeilinger (GHZ) states, preserving coherence across the entire array, or with $W$ states, collapsing the operation into pairwise readout~\citep{pra}. Although a classical Fourier transform of extracted pairwise visibilities may be performed, the GHZ approach enables a quantum Fourier transform directly on the stored quantum state (see Fig.~\ref{fig:network}). Coherent processing of the visibilities in the latter case results in an additional improvement in the signal-to-noise ratio, because the noise associated with pairwise measurements is avoided. The exact improvement depends on the nature of the source distribution, but it can be on the order of $\sqrt{N}$~\citep{pra}.
   
	In conclusion, we have proposed a protocol for performing nonlocal interferometry over a quantum network, which is relevant for astronomical imaging. By encoding the quantum state of the incoming photons into memory, we realize an effective ``event-ready'' scheme with efficient entanglement expenditure. The nonlocality is vital for removing vacuum noise in imaging weak thermal light, and distributed entanglement circumvents transmission losses. Hence, our scheme enables near-term quantum networks to serve as a platform for powerful optical interferometers that are demonstrably superior to what can be achieved classically. Furthermore, quantum algorithms can be used to process the stored signals such that the stellar intensity distribution can be inferred with a further improvement in the signal-to-noise ratio. 
	
	Although we focused on addressing fundamental limitations, real-world interferometry suffers from many other practical challenges such as stabilization and atmospheric phase fluctuations~\citep{Thompson2007}, which we address in Table~\ref{tab:error} of the Supplemental Material~\citep{SM}. For example, in Earth-based systems, atmospheric distortion can be tackled via a combination of adaptive optics and fringe tracking. Such techniques are already being deployed in smaller-scale astronomical interferometers~\cite{Stee2017} and are fully compatible with our proposal. Moreover, these challenges and control methods do not scale unfavorably with the baseline beyond certain physical correlation lengths~\citep{Thompson2007}, such that the construction of very large telescope arrays may be envisioned. Alternatively, space-based implementation avoids many of these technical issues, potentially extending over the $10^4$ km scale~\citep{Monnier2003} using entanglement distributed by satellites~\citep{Yin2017}. Because the astronomical origin of the weak thermal light is not crucial, such networks could also be applied to terrestrial imaging.
	
	\begin{acknowledgments}
	We thank A. Aspuru-Guzik, M. Bhaskar, F. Brand\~ao, M. Christandl, I. Chuang, I. Cirac, J. Stark, C. Stubbs, H. Zhou, and P. Zoller for illuminating discussions and useful comments. This work was supported by the National Science Foundation (NSF), the Center for Ultracold Atoms, the NSF Graduate Research Fellowship (E.T.K.), the Vannevar Bush Faculty Fellowship, ERC Grant Agreement No. 337603, the Danish Council for Independent Research (Sapere Aude), Qubiz -- Quantum Innovation Center, and VILLUM FONDEN via the QMATH Centre of Excellence (Grant No. 10059).
	\end{acknowledgments}
	
	\bibliography{prl}

	\clearpage	
	\onecolumngrid
	
	\begin{center}
		\textbf{Supplemental Material: Optical Interferometry with Quantum Networks}
	\end{center}
	
	In this supplement, we give the details of the transfer of the photonic density matrix to the memory qubits. Furthermore, we derive the Fisher information used to assess the parameter-estimating ability of our scheme in the presence of imperfections. In addition, we illustrate the optimality of the $\log 1/\epsilon$ scaling in resources via entropic quantities. By comparing the technical specifications of our protocol to the current state-of-the-art, we then show that near-term implementation is feasible. We close with a discussion of general challenges in astronomical interferometry and how they can be overcome in our scheme.
	
	\section{Encoding}
	
	We describe the encoding of the photonic state into the quantum memories in more detail.  As explained in the main text, a logical controlled not gate is first performed between the photonic degrees of freedom and the memory qubits, resulting in the following state:
	\begin{equation} \label{eq:supp1}
		\rho_{AB}\otimes\bar{\rho}_{0}\xrightarrow{\overline{\text{CX}}_m}(1-\epsilon)\rho_{\text{vac}}\otimes\bar{\rho}_{0}+\frac{\epsilon(1+\abs{g})}{2}\ket{\Psi^+_{\theta,m}}\bra{\Psi^+_{\theta,m}}+\frac{\epsilon(1-\abs{g})}{2}\ket{\Psi^-_{\theta,m}}\bra{\Psi^-_{\theta,m}}+O(\epsilon^2)\,,
	\end{equation}
	where $\bar{\rho}_{0}=\ket{\bar{0},\bar{0}}\bra{\bar{0},\bar{0}}$, $\ket{\Psi^{\pm}_{\theta,m}}=(\ket{0,1}\ket{\bar{0},\bar{1}_m}\pm e^{i\theta}\ket{1,0}\ket{\bar{1}_m,\bar{0}})/\sqrt{2}$, and $\ket{\bar{0},\bar{1}_m}$ denotes memory state $\ket{\bar{0}}$ at node $A$ and state $\ket{\bar{1}_m}$ at node $B$. Measuring the photonic degrees of freedom in the $X$ basis corresponds to projecting onto the photonic states $\ket{\pm}=\left(\ket{0}\pm\ket{1}\right)/\sqrt{2}$. Such a measurement will output one of four equally likely outcomes in the set $\{\ket{+,+},\ket{+,-},\ket{-,+},\ket{-,-}\}$. For each measurement outcome, the resulting state of the memories will be
	\begin{eqnarray}
		\{\ket{+,+},\ket{-,-}\} &:& (1-\epsilon)\bar{\rho}_{0}+\frac{\epsilon(1+\abs{g})}{2}\ket{\bar{\psi}^+_{m,\theta}}\bra{\bar{\psi}^+_{m,\theta}}+\frac{\epsilon(1-\abs{g})}{2}\ket{\bar{\psi}^-_{m,\theta}}\bra{\bar{\psi}^-_{m,\theta}}+O(\epsilon^2)\,, \\
		\{\ket{-,+},\ket{+,-}\} &:& (1-\epsilon)\bar{\rho}_{0}+\frac{\epsilon(1-\abs{g})}{2}\ket{\bar{\psi}^+_{m,\theta}}\bra{\bar{\psi}^+_{m,\theta}}+\frac{\epsilon(1+\abs{g})}{2}\ket{\bar{\psi}^-_{m,\theta}}\bra{\bar{\psi}^-_{m,\theta}}+O(\epsilon^2)\,,
	\end{eqnarray}
	where $\ket{\bar{\psi}^{\pm}_{m,\theta}}=\left(\ket{\bar{0},\bar{1}_m}\pm e^{i\theta}\ket{\bar{1}_m,\bar{0}}\right)/\sqrt{2}$. Depending on the measurement outcome, the memory state acquires a local phase of $-1$ compared to the original photonic state. In order to keep track of this phase, the photon's arrival time is needed, which is tracked efficiently in the code. 
	
	\section{Fisher information}
	
	The Fisher information $F$ describes the sensitivity of a measurement (positive operator-valued measure POVM $E(y)$, corresponding to outcome $y$) to the parameters $g_i$ of a quantum state (density matrix $\rho$). The first-order coherence $g=g_1+i g_2$ is a complex number that consists of two real parameters $g_1,g_2$ that are to be estimated, so the Fisher information becomes a matrix of rank two. From \cite{Tsang2011},
	\begin{equation}
		F=\sum_y \frac{D(y|g)}{P(y|g)}\,,\quad
		D(y|g)=\begin{pmatrix} (\frac{\partial P(y|g)}{\partial g_1})^2 & \frac{\partial P(y|g)}{\partial g_1} \frac{\partial P(y|g)}{\partial g_2} \\ \frac{\partial P(y|g)}{\partial g_2} \frac{\partial P(y|g)}{\partial g_1} & (\frac{\partial P(y|g)}{\partial g_2})^2 \end{pmatrix}\,,\quad
		P(y|g)=\tr(E(y)\rho)\,.
		\label{eq:fisher}
	\end{equation}
	
	First, consider the quantum state $\rho$ that is to be measured. The stellar light incident on two spatially separated sites $A,B$ is described by a bipartite thermal state~\citep{Tsang2011}:
	\begin{equation}
		\rho_{AB}=\int\,\mathrm{d}^2\alpha\,\mathrm{d}^2\beta\ket{\alpha,\beta}\bra{\alpha,\beta} \exp[-\begin{pmatrix} \alpha^* & \beta^* \end{pmatrix} \Gamma^{-1} \begin{pmatrix} \alpha \\ \beta \end{pmatrix}] / (\pi^2 \det\Gamma)\,,\qquad
		\Gamma=\begin{pmatrix}\langle a^\dagger a\rangle & \langle b^\dagger a\rangle \\ \langle a^\dagger b\rangle & \langle b^\dagger b\rangle \end{pmatrix} = \frac{\epsilon}{2}\begin{pmatrix} 1 & g \\ g^* & 1 \end{pmatrix}\,,
		\label{eq:thermal}
	\end{equation}
	where $\alpha,\beta$ are the coherent state amplitudes and $a,b$ are the annihilation operators of the optical modes at the two respective sites $A,B$. This state can be expanded in the Fock basis:
	\begin{eqnarray}
		\rho_{AB}&=&\frac{1}{1+\epsilon+\epsilon^2(1-|g|^2)/4}\rho^{(0)} \nonumber\\
		&+&\frac{\epsilon}{[1+\epsilon+\epsilon^2(1-|g|^2)/4]^2}\big[(1+\epsilon(1-|g|^2)/2)(\ket{0,1}\bra{0,1}+\ket{1,0}\bra{1,0}) + (g\ket{1,0}\bra{0,1}+h.c.)\big]/2 \nonumber\\
		&+&\frac{\epsilon^2[1+(2\epsilon-1)(1-|g|^2)/4+\epsilon^2((1-|g|^2)/4)^2]}{[1+\epsilon+\epsilon^2(1-|g|^2)/4]^2}\rho^{(>1)}\\
		&\xrightarrow{|g|=1}&\frac{1}{1+\epsilon}\rho^{(0)} + \frac{\epsilon}{(1+\epsilon)^2}\rho^{(1)}+\frac{\epsilon^2}{(1+\epsilon)^2}\rho^{(>1)}\,,
	\end{eqnarray} 
	where $\rho^{(0)}=\rho_\text{vac}$, $\rho^{(1)}=\ket{\psi^+_\theta}\bra{\psi^+_\theta}$, $\rho^{(>1)}$ contains multiple-photon events with $O(\epsilon^2)$; we have set $|g|=1$ to simplify the expression. This state is encoded into the memory in each of the $M$ time bins. The final decoded state $\rho$ in the readout qubit has coefficients given by a trinomial distribution of the input coefficients:
	\begin{equation}
		\rho_{AB}^{\otimes M}\to\rho = \Big[\frac{1}{1+\epsilon}\Big]^M\rho^{(0)}+M\Big[\frac{1}{1+\epsilon}\Big]^{M-1}\Big[\frac{\epsilon}{(1+\epsilon)^2}\Big]\rho^{(1)}+\Big(1-\frac{1}{(1+\epsilon)^M}-\frac{M\epsilon}{(1+\epsilon)^{M+1}}\Big)\rho^{(>1)}\,.
	\end{equation} 
	The vacuum component $\rho^{(0)}$ is post-selected out through the parity checks described in the main text. The multiple-photon component $\rho^{(>1)}$ is completely depolarized in our worst-case error model, where multiple photons in a single time bin and single photons spread out over multiple time bins result in uniformly random classical behavior of the readout qubit: $\rho^{(>1)}\to\rho_\text{mix}$, where $\rho_\text{mix}$ is the maximally mixed qubit state $I/2$. Then, the (subnormalized, post-selected) readout state is
	\begin{equation}
		\rho\to\frac{M\epsilon}{(1+\epsilon)^{M+1}}\rho^{(1)}+\Big(1-\frac{1}{(1+\epsilon)^M}-\frac{M\epsilon}{(1+\epsilon)^{M+1}}\Big)\rho_\text{mix}\,.
	\end{equation} 
	It can be rewritten in a convenient form where the success probability $p$ is the trace and the coherence $c$ is the magnitude of the off-diagonal component:
	\begin{equation}
		\rho=p(c\rho^{(1)}+(1-c)\rho_\text{mix})\,,\quad
		p = \frac{(1+\epsilon)^M-1}{(1+\epsilon)^M}\,,\quad
		c = \frac{M\epsilon}{[(1+\epsilon)^M-1](1+\epsilon)}\,.
		\label{eq:rho}
	\end{equation}
	The photonic state is mapped to memory as explained in the main text. All the one-qubit measurements, two-qubit gates, and entangled pairs used in the protocol will however have some infidelity. In our worst-case model, we describe all these errors as depolarizing channels acting on the state in Eq.~(\ref{eq:rho}): with probability $p_\text{ideal}$ the operation works ideally and the remaining $1-p_\text{ideal}$ of the time the state is lost, with uninformative measurement outcomes. \begin{equation}
		\rho\to p_\text{ideal}\rho+(1-p_\text{ideal})\rho_\text{mix}\,.
		\label{eq:depolarizing}
	\end{equation}
	An error event thus contributes directly to the maximally mixed component with probability $1-p_\text{ideal}$ and we can interpret $p_\text{ideal}$ as the coherent fraction $c$. The fidelity $f$ between an input pure state $\ket{\psi}$ and the channel output is
	\begin{equation} \label{eq:f(c)}
		f(\ket{\psi},c\ket{\psi}\bra{\psi}+(1-c)\rho_\text{mix})=(1+c)/2\,,
	\end{equation}
	which defines the fidelity of an operation. Thus, unideal operations contribute to the mixed component of the quantum state. They are enumerated as follows:
	\begin{enumerate}
		\item Readin (per telescope site)
		\begin{enumerate}
			\item \textbf{State transfer} from optical mode to auxiliary memory qubit
			\item \textbf{Encoding:} $\log_2(M+1)/2$ (on average) nontrivial two-qubit gates
			\item \textbf{Decoupling:} 1 relevant one-qubit measurement of the auxiliary qubit
		\end{enumerate}
		\item Readout (per pair of telescope sites)
		\begin{enumerate}
			\item \textbf{Decoding:} $\log_2(M+1)/2$ entangled pairs, $\log_2(M+1)$ two-qubit gates, and $\log_2(M+1)$ one-qubit measurements
			\item \textbf{Interference:} $\log_2(M+1)$ one-qubit measurements
		\end{enumerate}	
	\end{enumerate}
	Overall, for two-site operation, there are 2 state transfers (with success probability $p_t$ and fidelity $f_t$), $2(1+\log_2(M+1))$ one-qubit measurements (with fidelity $f_1$), $2\log_2(M+1)$ two-qubit gates (with fidelity $f_2$), and $\log_2(M+1)/2$ entangled pairs (with fidelity $f_e$). These operations modify the $p,c$ in Eq.~\ref{eq:rho} to
	\begin{equation}\label{eq:gateMod}
		p\to p \cdot p_t^2 \quad
		c\to c \cdot (2f_t-1)^2 (2f_1-1)^{2(1+\log_2(M+1))} (2f_2-1)^{2\log_2(M+1)} (2f_e-1)^{\log_2(M+1)/2}\,.
	\end{equation}
	
	Second, consider the measurement $E(y)$. The readout qubit is measured in a basis with tunable phase $\delta$; equivalently, a rotation $U_\delta=\ket{0}\bra{0}+e^{i\delta}\ket{1}\bra{1}$ and Hadamard gate $U_H$ are applied on the state before measurement in the computational basis. The outcome probability is given by $P(y|g)=\bra{y}U_H U_\delta\rho U_\delta^\dagger U_H^\dagger\ket{y}$ where $\ket{y}\in\{\ket{0},\ket{1}\}$. Hence, the POVM is $E(y)=U_\delta^\dagger U_H^\dagger\ket{y}\bra{y}U_H U_\delta$, which evaluates to
	\begin{eqnarray}
		E(0)&=&(\ket{0}+e^{-i\delta}\ket{1})(\bra{0}+e^{i\delta}\bra{1})/2 \label{eq:POVM0}\,,\\
		E(1)&=&(\ket{0}-e^{-i\delta}\ket{1})(\bra{0}-e^{i\delta}\bra{1})/2 \label{eq:POVM1}\,.
	\end{eqnarray}
	
	Now, we can calculate the Fisher information, given the state of Eqs.~\eqref{eq:rho},\eqref{eq:gateMod} and measurements of Eqs.~\eqref{eq:POVM0},\eqref{eq:POVM1}. We obtain 
	\begin{eqnarray}
		P(0|g)&=&p(1+c(g_1\cos\delta+g_2\sin\delta))/2\,,\\ P(1|g)&=&p(1-c(g_1\cos\delta+g_2\sin\delta))/2\,.
	\end{eqnarray}
	Then, the Fisher information matrix is
	\begin{equation}
		F=\frac{p}{(1/c^2)-\text{Re}^2(ge^{-i\delta})}
		\begin{pmatrix} \cos^2\delta & \sin\delta\cos\delta \\ \sin\delta\cos\delta & \sin^2\delta \end{pmatrix}\,,
	\end{equation}
	with trace norm
	\begin{equation}
		\lVert F\rVert=\frac{p}{(1/c^2)-\text{Re}^2(ge^{-i\delta})}\,.
	\end{equation}
	$\delta$ is a controllable phase that is swept over measurements to estimate the two quadratures of $g$. We set $\text{Re}(ge^{-i\delta})=0$ to obtain a lower bound: 
	\begin{equation}
		\lVert F\rVert\geq p c^2 = \frac{(M\epsilon)^2}{[(1+\epsilon)^M-1](1+\epsilon)^{M+2}} p_t^2 (2f_t-1)^4 (2f_1-1)^{4(1+\log_2(M+1))} (2f_2-1)^{4\log_2(M+1)} (2f_e-1)^{\log_2(M+1)}\,.
	\end{equation}
	
	\section{Minimum entanglement consumption}
	
	As described in the main text, the stellar light that hits the two sites $A,B$ per time bin is specified by a density matrix 
	\begin{equation}
		\rho_{AB}=(1-\epsilon)\rho_{\text{vac}}+\frac{\epsilon(1+\abs{g})}{2}\ket{\psi^{+}_{\theta}}\bra{\psi^+_{\theta}}+\frac{\epsilon(1-\abs{g})}{2}\ket{\psi^{-}_{\theta}}\bra{\psi^-_{\theta}}+O(\epsilon^2)\,,
	\end{equation}
	to first order in $\epsilon$, where $\ket{\psi^{\pm}_{\theta}}=(\ket{0,1}\pm e^{i\theta}\ket{1,0})/\sqrt{2}$ and $\rho_{\text{vac}}=\ket{0,0}\bra{0,0}$. Estimating $g=\lvert g \rvert e^{i\theta}$ optimally requires nonlocal measurements, in order to distinguish the single photon $\ket{\psi^{\pm}_{\theta}}$ against the vacuum $\rho_{\text{vac}}$~\citep{Tsang2011}, which may be expanded in the basis of $\ket{\phi^{\pm}}=(\ket{0,0}\pm\ket{1,1})/\sqrt{2}$. Measurements over the Bell basis use one maximally entangled pair (ebit) each. Nonetheless, the Bell states may be distilled to an information-theoretic limit to consume the minimum amount of entanglement.
	
	In our protocol, an ensemble $\rho_{AB}^{\otimes M}$ of incoming states is encoded sequentially in the quantum memories in blocks, which are processed nonlocally to distill an entangled pair with associated arrival time information. The consumption of ebits increases as $\log(1/\epsilon)$ since the number of time bins per block is $M\sim1/\epsilon$. However, the actual uncertainty of the joint state $\rho_{AB}$ from the perspective of one site $B$ is captured by the conditional entropy,
	\begin{eqnarray}
		S(A|B)_\rho&=&S(AB)_\rho-S(B)_\rho \,, \\
		S(B)_\rho&=&-\tr\rho_{B}\log\rho_{B} \,, \\
		\rho_{B}&=&\tr_A \rho_{AB} \,.
	\end{eqnarray}
	Operationally, the conditional entropy corresponds to the entanglement cost of state merging: Refs.~\citep{Horodecki2005,Horodecki2007} show that the quantum information from $A$ can, in principle, be transferred to $B$ using $S(A|B)_\rho$ ebits per copy of $\rho_{AB}$ together with local operations and classical communication. This result applies in the asymptotic regime (number of samples $1/\epsilon\to\infty$) and is optimal. All information about the parameter $g$ can subsequently be extracted locally. Hence, in our setting the theoretical limit to entanglement consumption is $\frac{1}{\epsilon}S(A|B)_\rho$ ebits. 
	
	Evaluating the entropies for our $\rho_{AB}$ in the weak-source limit $\epsilon\ll1$,
	\begin{eqnarray}
		S(B)_\rho&\to&\frac{\epsilon}{2}\log\frac{2}{\epsilon}\,,\\
		S(AB)_\rho&\to&\epsilon\log\frac{1}{\epsilon}\,,\\
		S(A|B)_\rho&\to&\frac{\epsilon}{2}\log\frac{1}{\epsilon}\,.
	\end{eqnarray}
	Then, the minimum entanglement consumption is, up to an $o(1)$ term from finite block length,
	\begin{equation}
		\frac{1}{\epsilon}S(A|B)_\rho\xrightarrow{\epsilon\ll 1}\frac{1}{2}\log\frac{1}{\epsilon}\,.
	\end{equation}
	This lower bound matches the scaling of our protocol, within a factor of 2, and is derived in the asymptotic limit for general local operations. We believe that our sequential compression of the incoming states may be the reason for the gap. Protocols obtaining the information-theoretic limit generally require all $M$ states to be stored such that quantum operations can be performed across the entire ensemble to distill the arrival time information~\citep{Ahn2006}. Meanwhile, the amount of classical communication in our protocol matches the limit of $\log(1/\epsilon)$ bits~\footnote{The rate is given by the mutual information $I(A;R)_\varphi$, where $\ket{\varphi}_{ABR}$ is the purification of $\rho_{AB}$, which in our case can be simplified to $S(AB)_\rho$.}. These bounds hold even though the state is not fully specified because $g$ is unknown~\citep{Bjelakovic2013}. We therefore conclude that our practical protocol is nearly optimal in terms of communication resources.
	
	\section{Technical specifications}
	
	The most demanding requirement of our protocol is the encoding operation, which occurs over an interval set by the detector bandwidth $\delta_f$. Otherwise, if the gate time $\tau_g>1/\delta_f$, additional qubits may be introduced to operate in parallel~\citep{pra}. The decoding operation is performed at the more lenient photon arrival rate $\delta_f\epsilon$, which also sets the minimum lifetime of the memories. For shorter coherence times $\tau_m<1/(\delta_f\epsilon)$, the memories are read out more frequently. Under these experimental constraints, computational resources scale as $(\tau_g\delta_f)\log_2(\tau_m\delta_f)$ and entanglement consumption as $1/\tau_m\log_2(\tau_m\tau_g\delta_f^2)$. These solutions artificially improve $\tau_m$ and $\tau_g$, and do not adversely affect the error accumulation discussed above; i.e., it remains logarithmic. In our example reflecting realistic optical interferometry, the detector bandwidth is 10 GHz (100 ps) and the photon arrival rate is 1 kHz (1 ms). Our memory-based scheme, with decoding rate decreased to avoid errors and maximize information extraction, reduces the demand on the quantum network to provide entangled pairs at a rate of 200 kHz (5 $\mu$s),
	
	Storing information in quantum memory is advantageous if the quantum resources are not too noisy. In our worst-case error model, memoryless scaling is observed when the error parameter $\nu\equiv(2f_1-1)^4(2f_2-1)^4(2f_e-1)$ drops to $\sim$60\%. Then, the fidelity of quantum operations (assuming for simplicity that the one-qubit measurement fidelity $f_1$ equals the two-qubit gate fidelity $f_2$) should be at least 97\%. We assume the trivial CX instances have error rate less than the photon arrival rate. Similarly, the minimum fidelity of entangled pairs $f_e$ is 80\%. In general, there is leeway on the amount and source of noise, which may be relaxed for a given error model and source intensity $\epsilon$.
	
	We compare to the capabilities of current technology in Table~\ref{tab:specifications}. We see that most components are close to the required level for three state-of-the-art cavity QED platforms: atoms, quantum dots, and SiV centers in diamond. Atoms and SiV qubits benefit from longer coherence times, but also have slower gate times compared to quantum dots. These hardware characteristics may become advantageous in different operating regimes of the telescope: e.g., for weaker sources with smaller $\epsilon$. Photon detection-based state transfer~\citep{pra} alleviates the bottleneck associated with slow spin readout; photon detectors can operate on the scale of picoseconds~\citep{Waks2002}. Photon-mediated gates also are quick and moreover do not cause a nontrivial CX in the absence of a control photon, as desired. Hence, the encoding may be achieved with fast photonic operations. For qubit-qubit gates, optical cavities enable integrated error detection so that any trivial CX errors can be postselected out~\citep{Borregaard2015}. The outstanding challenge is the capabilities of present quantum networks, themselves. Nonetheless, using the most recent light-collection techniques, entanglement distribution rates can be improved to 130 kHz~\citep{Stockill2017}. We hope that our application further motivates such development of quantum networks.
	
	\begingroup
	\squeezetable
	\begin{table*} [htbp!]
		\begin{tabular}{*{9}{c|} c }
			& \multicolumn{2}{c|}{One-qubit gate} & \multicolumn{2}{c|}{Two-qubit gate} & \multicolumn{2}{c|}{Atom-photon gate} & Memory & \multicolumn{2}{c}{Readout} \\ 
			Experimental platform & $f$ & $\tau$ & $f$ & $\tau$ & $f$ & $\tau$ & $\tau$ & $f$ & $\tau$ \\ \hline\hline
			Rb atoms in cavity~\citep{Reiserer2014,PhysRevX.8.011018,korber2018decoherence} & 95\% & 7.5 $\mu$s & 92\% & 0.9 $\mu$s & 87\% & 0.7 $\mu$s & 100 ms & 99.65\% & 3 $\mu$s \\ \hline
			Quantum dots \& nanophotonics~\citep{Stockill2017,Kim2010,PhysRevB.97.241413,Sun2018} & 98\% & 3 ps & 80\% & 150 ps & 78\% (99\%)$^1$ & 75 ps & 4 $\mu$s & 98\%$^2$ & 75 ns \\ \hline
			SiV in diamond nanocavity~\citep{Sukachev2017,Evans662} & 99\% & 10 ns & 92\%$^3$ & 10 ns & 87\%$^3$ & 5 ns$^3$ & 13 ms (1 s)$^4$ & 97\% & 100 ns$^2$ \\
		\end{tabular}
		\caption{Accounting of the performance of quantum technology that is particularly suitable for implementation of our scheme, in terms of fidelity $f$ and timescale $\tau$. $^1$Expected spin-photon entanglement fidelity for hole-spins~\citep{korber2018decoherence}. $^2$Assuming modest 10\% collection efficiency. $^3$Using SiV-photon cooperativity $C = 23$ demonstrated in \citep{Evans662}, which is on par with that of the Rb-cavity system used to demonstrate atom-photon gates~\citep{Reiserer2014} and two-qubit gates~\citep{PhysRevX.8.011018}. $^4$Using proximal $\mathrm{^{13}C}$ nucleus~\citep{Sukachev2017,Maurer1283}.
			\label{tab:specifications}} 
	\end{table*}
	\endgroup
	
	\section{Error budget for Astronomical Interferometry}
	
\begin{figure} [h]
		\centering
		\includegraphics[width=\textwidth]{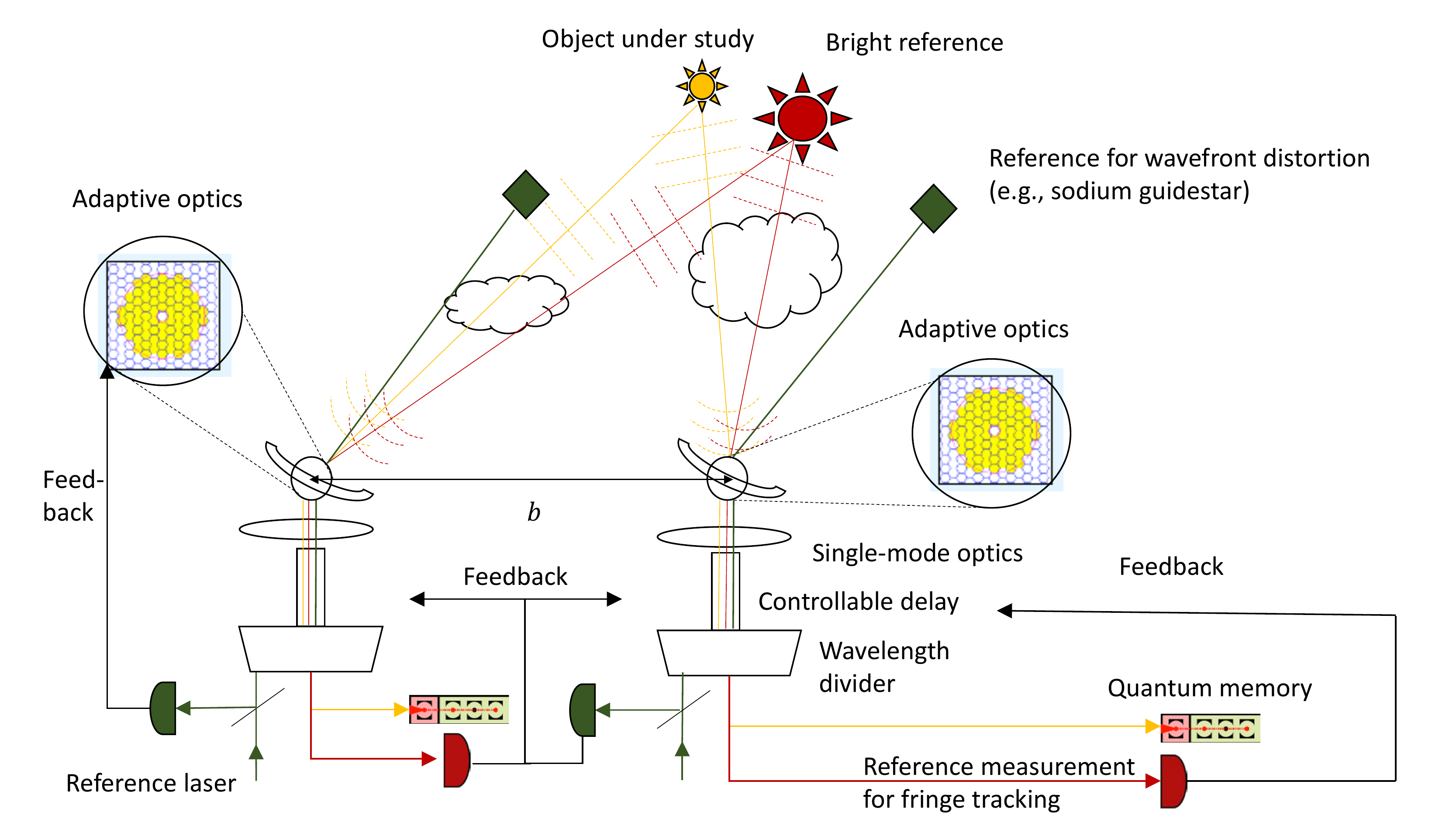}
		\caption{Overview of a basic scheme for stabilization. In the first feedback loop, light from a sodium guidestar is used in combination with adaptive optics and single-mode coupling to correct wavefront fluctuations due to atmospheric disturbance. Then, a wavelength-division multiplexer splits the light of a bright reference from that of the astronomical object under study. Through fringe tracking of the bright reference, fed back onto variable delay on the common path, differential path fluctuations can be corrected. \label{fig:errors}}
	\end{figure}

	Telescope arrays that rely on direct interference of light require interferometric stability for proper operation. Several recent experiments on the CHARA and VLTI telescope arrays in the visible and near-IR domain tackle stabilization issues via hierarchical levels of control, combining adaptive optics and differential measurement (``fringe tracking'') techniques as well as single-mode optics~\cite{Stee2017,gravity,CHARAerror}. Similar techniques, using a hierarchy of feedback loops (see Fig.~\ref{fig:errors}), could be used to tackle the main sources of instability, which can be roughly divided as follows:	
	\begin{itemize}
		\item Fluctuations in the optical path length difference between the arms of the interferometer, directly affecting the interference. Required stability $\ll\lambda$.
		\item Fluctuations in the effective baseline $b$ of the interferometer, affecting the angular resolution $\delta \phi$. Required stability $\frac{\delta b}{b}\ll\frac{\delta\phi}{\phi}$.
	\end{itemize}

	The maximum angular resolution of current telescope arrays working in the optical or infrared regime is on the scale of milliarcseconds (mas). Going to sub-mas resolution opens up many new scientific opportunities~\citep{Stee2017}. For imaging in the optical domain around 550 nm, baselines on the order of 1--10 km are necessary to reach resolutions of 100--10 $\mu$as. In Table~\ref{tab:error}, the dominant error sources for such an optical telescope array are summarized together with their tolerances. Potential methods for suppressing the errors to these levels are described in the last column. Note that some of the requirements for interferometric stability are relaxed because of the entanglement-assisted nature of our proposal. Instead of requiring the optical path between sites of the interferometer to be stable on $\ll\lambda$ scales (0.1 $\mu$m), the \emph{entangled pairs} need to be phase-stable to prevent loss of visibility. The creation of phase-stable entangled memory pairs using two-photon interference schemes has recently been demonstrated experimentally~\cite{Bernien2013}, with sensitivity to fluctuations on the scale of $c/\nu$, where $c$ is the speed of light and $\nu$ is the qubit energy separation. For typical spin qubits, the gap is on the order of gigahertz corresponding to centimeter-scale stabilization.
	
	Consider realizing long-baseline interferometry with conventional direct detection. Fiber coupling at visible wavelengths would be infeasible already for baselines around $\sim$1 km since attenuation losses are on the order of 20--30 dB/km. Wavelength conversion to the telecom band would significantly decrease fiber losses to around 0.2 dB/km, pushing baselines to $\sim$10 km assuming perfect conversion. However, obtaining low-noise and efficient conversion is a major difficulty. Finally, consider free-space, non-vacuum links. For clear weather conditions~\cite{Kaushal2017}, losses are on the order of 0.5 dB/km, which quickly increase for non-clear conditions along with significant beam errors. Meanwhile, our quantum network protocol makes use of pre-shared entangled pairs. These pairs are distributed using quantum repeater techniques~\citep{Briegel1998} that in principle have no distance limit, realistically allowing for long-baseline interferometry on the scale of 100--1000s of kilometers~\citep{Muralidharan2016}, infeasible with direct interferometry.
	
	\begingroup
	\squeezetable
	\begin{table*} [h]
		\begin{tabular}{ c | p{5 cm} | c | p{5 cm} }
			Error & Description & Tolerance & Correction \\ \hline \hline
			Atmospheric turbulence & These irregularities distort the optical wavefronts (1) and change the optical path length difference (2). The wavefront distortion can be transformed into light intensity fluctuations via single-mode fiber coupling~\cite{Stee2017}. Uncorrelated fluctuations between sites introduce errors in the differential optical path length. Without correction, these \emph{global} and \emph{differential} errors limit the resolution to arcseconds and largely wash out the interference. & 0.1 $\mu$m & Adaptive optics and single-mode fiber coupling mitigate the effect of wavefront distortion (1). Differential path length fluctuations (2) are corrected using fringe tracking on bright reference stars in the vicinity of the astronomical object under study~\cite{Stee2017}. For GRAVITY, the difference in optical path length is reduced to $\sim$10 nm with current adaptive optics and fringe tracking.  \\ \hline
			Tip-tilt & Any or all of the telescopes in the array are pointed inaccurately. Global errors result in incorrect estimation of angular position (see also stellar orientation error below). Differential errors change the optical path length. & 0.1 $\mu$m & These errors are tackled using telescope encoders in combination with fringe tracking of a bright reference star. VLTI reaches angular resolutions of 10 mas at wavelengths of 1--2 $\mu$m, requiring stabilization of $\sim$1 $\mu$m over 100 m.  \\ \hline
			Stellar orientation & Error in tracking the (continuously moving) relative position of the stellar object with respect to the Earth. Differential errors change the optical path length. & 0.1 $\mu$m & Differential errors are mitigated via fringe tracking of bright reference stars nearby the object under study. For global errors, the dominant contribution to incorrect tracking for GRAVITY is the uncertainty in the time-stamp of the data due to finite integration time. A timing error of 1/15 s over a 100 m baseline leads to an effective error of 0.5 mm. This error may be mitigated in our proposal, which keeps track of photon arrival time. \\ \hline
			Baseline & Variations in the distance between telescope sites. Depending on the entanglement generation scheme, these errors may not affect the interference itself but can influence the visibility through an effective angular fluctuation. & 1 cm & The distance can be tracked via reference lasers. Our quantum-assisted proposal does not require beam propagation of the stellar light between sites. Assuming entanglement generation schemes that are only second-order sensitive to length fluctuations~\cite{Bernien2013}, the stability requirement is set by the energy separation of the qubit levels, typically in the gigahertz range. Therefore, the required stability is largely determined by the tolerable angular fluctuation (100--10 $\mu$as). 
		\end{tabular}
		\caption{Description of the dominant errors for long-baseline interferometry with quantum networks. The state-of-the-art quoted in the last column is based on specifications of the VLTI astronomical interferometer~\citep{VLTI} and its GRAVITY instrument~\citep{gravity}. \label{tab:error}} 
	\end{table*}
	\endgroup

\end{document}